\documentclass[aps,twocolumn,floatfix,10pt,superscriptaddress,prl]{revtex4-1}

\usepackage{bm}
\usepackage{graphicx}
\usepackage{subfigure}
\usepackage{epstopdf}
\usepackage{siunitx}

\usepackage{color}

\def\bem#1{\begin{mathletters}\label{#1}}
\def\eml{\end{mathletters}}

\def\4#1{{\boldsymbol{#1}}}
\def\8#1{{\widetilde{#1}}}

\begin{document}

\title {Negative charge enhancement of near-surface nitrogen vacancy centers by multicolor excitation}

\author{I. Meirzada}
\affiliation{The Racah Institute of Physics, The Hebrew University of Jerusalem, Jerusalem 91904, Israel}

\author{Y. Hovav}
\affiliation{Dept. of Applied Physics, Rachel and Selim School of Engineering, The Hebrew University of Jerusalem 91904, Israel}

\author{S. A. Wolf}
\affiliation{The Racah Institute of Physics, The Hebrew University of Jerusalem, Jerusalem 91904, Israel}
\affiliation{The Center for Nanoscience and Nanotechnology, The Hebrew University of Jerusalem, Jerusalem 91904, Israel}

\author{N. Bar-Gill}
\email{bargill@phys.huji.ac.il}
\thanks{Corresponding author.}
\affiliation{Dept. of Applied Physics, Rachel and Selim School of Engineering, Hebrew University, Jerusalem 91904, Israel}
\affiliation{The Racah Institute of Physics, The Hebrew University of Jerusalem, Jerusalem 91904, Israel}
\affiliation{The Center for Nanoscience and Nanotechnology, The Hebrew University of Jerusalem, Jerusalem 91904, Israel}

\begin{abstract}
Nitrogen-Vacancy (NV) centers in diamond have been identified over the past few years as promising systems for a variety of applications, ranging from quantum information science to magnetic sensing. This relies on the unique optical and spin properties of the negatively charged NV. Many of these applications require shallow NV centers, i.e. NVs that are close (a few nm) to the diamond surface. In recent years there has been increasing interest in understanding the spin and charge dynamics of NV centers under various illumination conditions, specifically under infra-red (IR) excitation, which has been demonstrated to have significant impact on the NV centers’ emission and charge state. Nevertheless, a full understanding of all experimental data is still lacking, with further complications arising from potential differences between the photo-dynamics of bulk and shallow NVs. Here we suggest a generalized quantitative model for NV center spin and charge state dynamics under both green and IR excitation. We experimentally extract the relevant transition rates, providing a comprehensive model which reconciles all existing results in the literature. Moreover, we identify key differences between the photo-dynamics of bulk and shallow NVs, and use them to significantly enhance the initialization fidelity of shallow NVs to the useful negatively-charged state. 


\end{abstract}

\maketitle

The nitrogen-vacancy (NV) center \cite{doherty_nitrogen-vacancy_2013} has attracted significant attention over the past several years, as a model quantum system for a wide range of applications, such as quantum information processing \cite{Awschalom_memory} and quantum sensing \cite{taylor_high-sensitivity_2008,Loretz, clevenson_broadband_2015, trusheim_wide-field_2016,dolde_electric-field_2011}. In addition, shallow NVs positioned a few nm from the diamond surface can be used to detect nuclear or electron spins outside the diamond \cite{staudacher_nuclear_2013,mamin_nanoscale_2013}. 
Interest in the NV platform stems from the ability to readout and initialize its spin state optically, due to the NV photo-dynamics, mainly under green (532nm) laser excitation. 
In addition, NVs show relatively long coherence times \cite{bar-gill_solid-state_2013,farfurnik_optimizing_2015} even at room temperature \cite{childress_coherent_2006,robledo_high-fidelity_2011}, with actual values depending on the NV surrounding, including its distance from the surface \cite{romach_spectroscopy_2015,ohno_engineering_2012,WangCoherence2016}. These properties appear in the negative charge state (NV$^-$), and are absent in its neutral charge state (NV$^0$). However, charge state manipulation was shown to be a useful tool for various applications, such as spin readout enhancement \cite{Shields_ChargeStateReadout_2015} and quantum computation \cite{jayakumar_OpticalPatterning_2016}. 
 

It was recently shown that a combined green and IR (1064 nm) excitation has a dramatic impact on the NV's emission \cite{geiselmann_fast_2013, hopper_near-infrared-assisted_2016,Lai_2013} and charge state \cite{ji_charge_2016}. 
These works, together with earlier studies of NV charge dynamics \cite{aslam_photo-induced_2013,manson_nitrogen-vacancy_2006,robledo_spin_2011}, suggest various models for describing specific experimental results. However, these proposed models do not fully explain all of the existing literature, and cannot reproduce the measured dynamics. 
Furthermore, a comparison between the behavior of bulk vs. shallow NVs is currently missing. 

In this work we study experimentally the photo-dynamics of NV centers due to excitations with both green (532 nm) and IR (1064 nm) light, for a wide range of laser powers. The IR wavelength was chosen to avoid ionization from the singlet states, as reported in \cite{hopper_near-infrared-assisted_2016}. 
We focus on shallow NVs 
and compare to bulk NVs in a high-pressure high-temperature (HPHT) sample. We construct a model that can explain these results, and extract experimentally the relevant rates/cross-sections. We show that this model is consistent with both the data presented here and previously published results, suggesting a full understanding of NV photo-dynamics under the studied conditions. Finally, we identify a significant difference between the charge dynamics of bulk and shallow NVs, as well as an optimized two-color initialization procedure for enhancing NV$^-$ population of shallow NVs. 


The measured results that are shown below were performed using a homebuilt confocal microscope. Green (532 nm) and IR (1064 nm) continuous wave lasers were focused into a diffraction limited spot using an oil immersion objective. Fluorescence (PL) was collected from the same objective, and directed into a single photon counter. Single shallow NVs were measured in a high purity chemical vapor deposition (CVD) sample, as well as in an HPHT sample, in which bulk NVs were also measured. 
We use narrow band-pass filters to quantify the charge-state population change (NV$^-$ vs. NV$^0$) through the filtered PL level (see \cite{SuppMat,aslam_photo-induced_2013}). 



\begin{figure}[tbh]
\subfigure[]{
\includegraphics[trim = 0mm 1mm 0mm 3mm, clip, width=0.48 \linewidth]{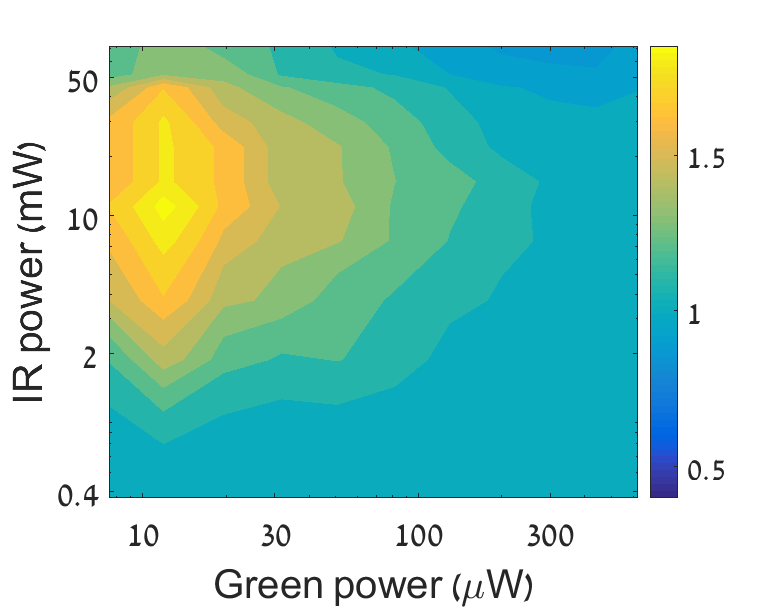}}
\subfigure[]{
\includegraphics[trim = 0mm 1mm 0mm 3mm, clip, width=0.48 \linewidth]{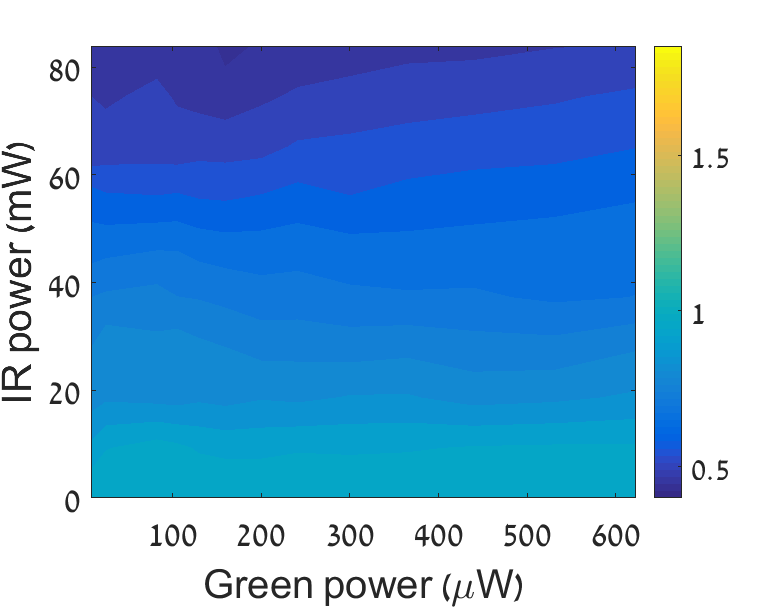}}

\protect\caption{IR excitation impact on Steady-state fluorescence of shallow and bulk NVs. (a) shallow NV$^-$ steady-state fluorescence under simultaneous green and IR excitations as a function of both green and IR powers on a logarithmic scale, normalized by the fluorescence without the IR excitation. Regions of enhancement and suppression of the fluorescence can be observed. (b) bulk NV$^-$ steady-state fluorescence under simultaneous green and IR excitations as a function of both green and IR powers on a logarithmic scale, normalized by the fluorescence without the IR excitation. Only suppression of the fluorescence is observed.}
\label{fig:steadystate}
\end{figure}

We first measured the steady-state NV$^-$ PL under green illumination, and its change due to the addition of IR excitation for surface NVs in both HPHT and high purity CVD samples, and compared it to bulk NVs in an HPHT sample. 
Fig. \ref{fig:steadystate} 
shows this change in PL as a function of both IR and green excitation powers. Unlike previous reports which exhibited either enhancement \cite{ji_charge_2016} or suppression \cite{geiselmann_fast_2013} in steady-state PL due to 1064 nm IR illumination, for the shallow NVs [Fig. \ref{fig:steadystate}(a)] we observe a smooth transition between these effects, with a non-monotonous behavior as a function of both the green and IR power for the same NV center (these results have been observed for several different NVs in this sample), in agreement with \cite{hopper_near-infrared-assisted_2016}. 
This suggests that a single mechanism underlies both enhancement and suppression effects, which we now elucidate through dynamic (pulsed) experiments. 
However, for bulk NVs [Fig. \ref{fig:steadystate}(b)] we find significantly different behavior. the green laser power has very little effect on the fluorescence from bulk NVs (in the range of powers we used). furthermore, as opposed to the non-monotonous and logarithmic dependence in shallow NVs, in bulk NVs the steady state fluorescence ratio changed monotonically and linearly with IR power. 
Moreover, for bulk NVs the suppression of NV$^-$ PL is more dominant compared to shallow NVs (more closely resembling the results of \cite{geiselmann_fast_2013}).%

\begin{figure}[tbh]
\subfigure[]{
\includegraphics[trim = 3mm 1mm 0mm 3mm, clip, width=0.46 \linewidth]{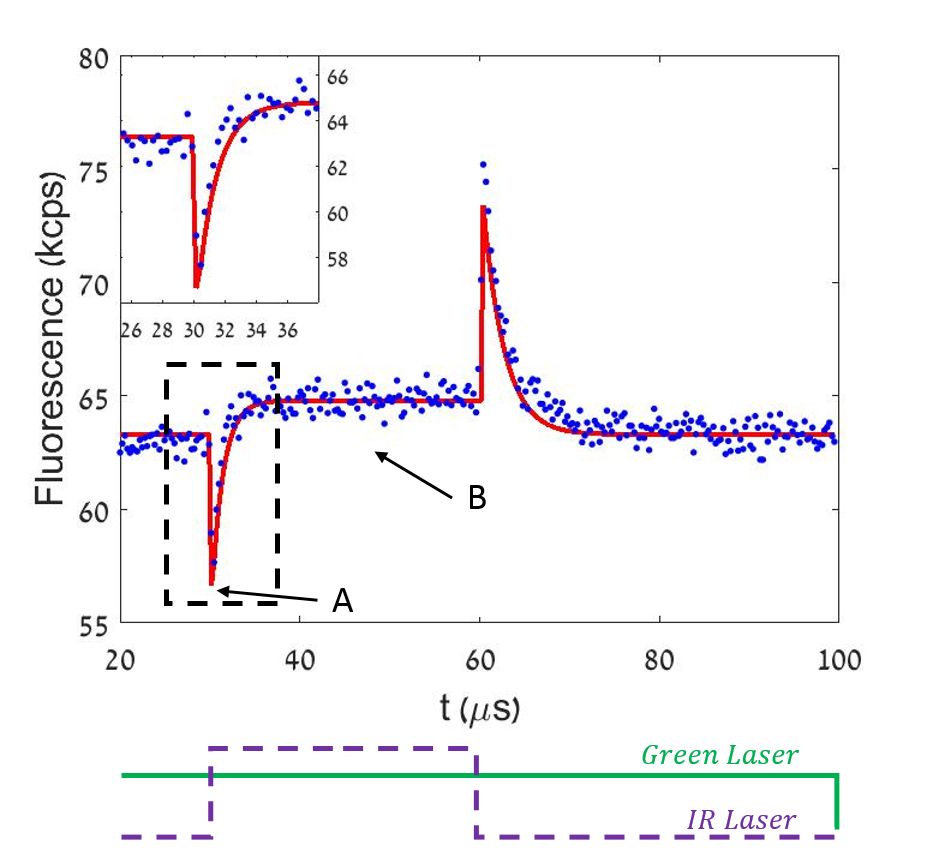}}
\subfigure[]{
\includegraphics[trim = 3mm 1mm 0mm 3mm, clip, width=0.46 \linewidth]{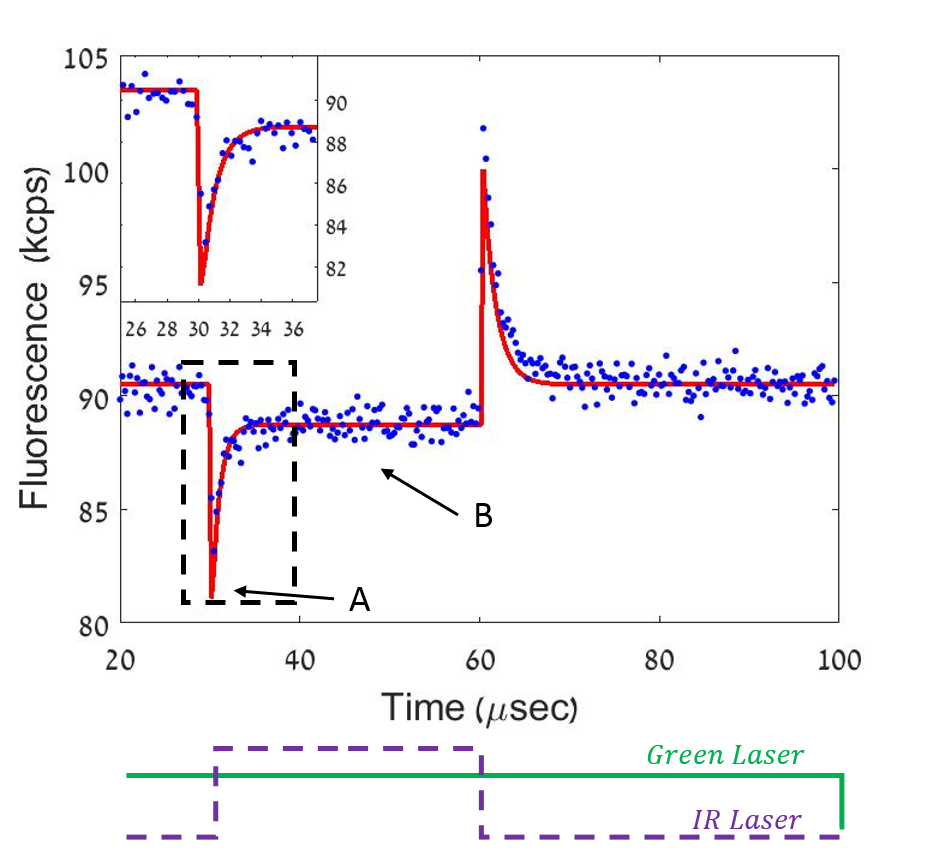}}

\protect\caption{IR excitation impact on NV$^-$ fluorescence as a function of time. Blue dots - experimental data, red curve - fit using numerical simulation according to the model presented in Fig.  \ref{fig:energyleveldiagram} using the rates extracted in Fig. \ref{fig:quenchIR}(a) Time resolved NV$^-$ fluorescence for 159 $\mu W$ green laser and 38 mW IR laser. A sharp suppression (A and inset) followed by a slow increase ends with an increase of steady state fluorescence (B). (b) Time resolved fluorescence for 241 $\mu W$ green laser and 38 mW IR laser. A sharp suppression (A and inset) followed by a slow increase ends with a decrease of steady state fluorescence (B).}
\label{fig:timeresolved}
\end{figure}

We next consider the photo-dynamics of shallow NVs, 
as shown in Fig. \ref{fig:timeresolved}, in analogy to the experiments performed in Ref \cite{ji_charge_2016} (see \cite{SuppMat}). 
These measurements were performed with a wide range of green and IR powers.

%

In Fig. \ref{fig:timeresolved} we plot the dynamics of two representative cases, with either (a) 159 $\mu$W or (b) 241 $\mu$W of green laser power, where in both cases 38 mW IR power was used. The former shows an enhancement of the steady-state NV$^-$ PL in the presence of IR illumination (B in Fig. \ref{fig:timeresolved}a) whereas the latter shows suppressed PL (B in Fig. \ref{fig:timeresolved}b). In both cases the IR resulting dynamics show two dominant timescales: a very fast (\textless 30 ns) quench (increase) in the PL as the IR excitation is turned on (off), followed by slower relaxation to steady-state (on a $\mu$s timescale) in the opposite direction. This dynamical picture suggests different, competing processes, which act on different timescales, such that the interplay between them dictates the steady-state result (enhancement or suppression of PL). The timescale of the fast quench dynamics observed in Fig. \ref{fig:timeresolved} seems independent of green excitation power. This suggests that the relevant processes are related to the excited state, namely single-photon ionization/recombination (I/R) from it. 

\begin{figure}[tbh]
{\includegraphics[width=0.85 \linewidth]{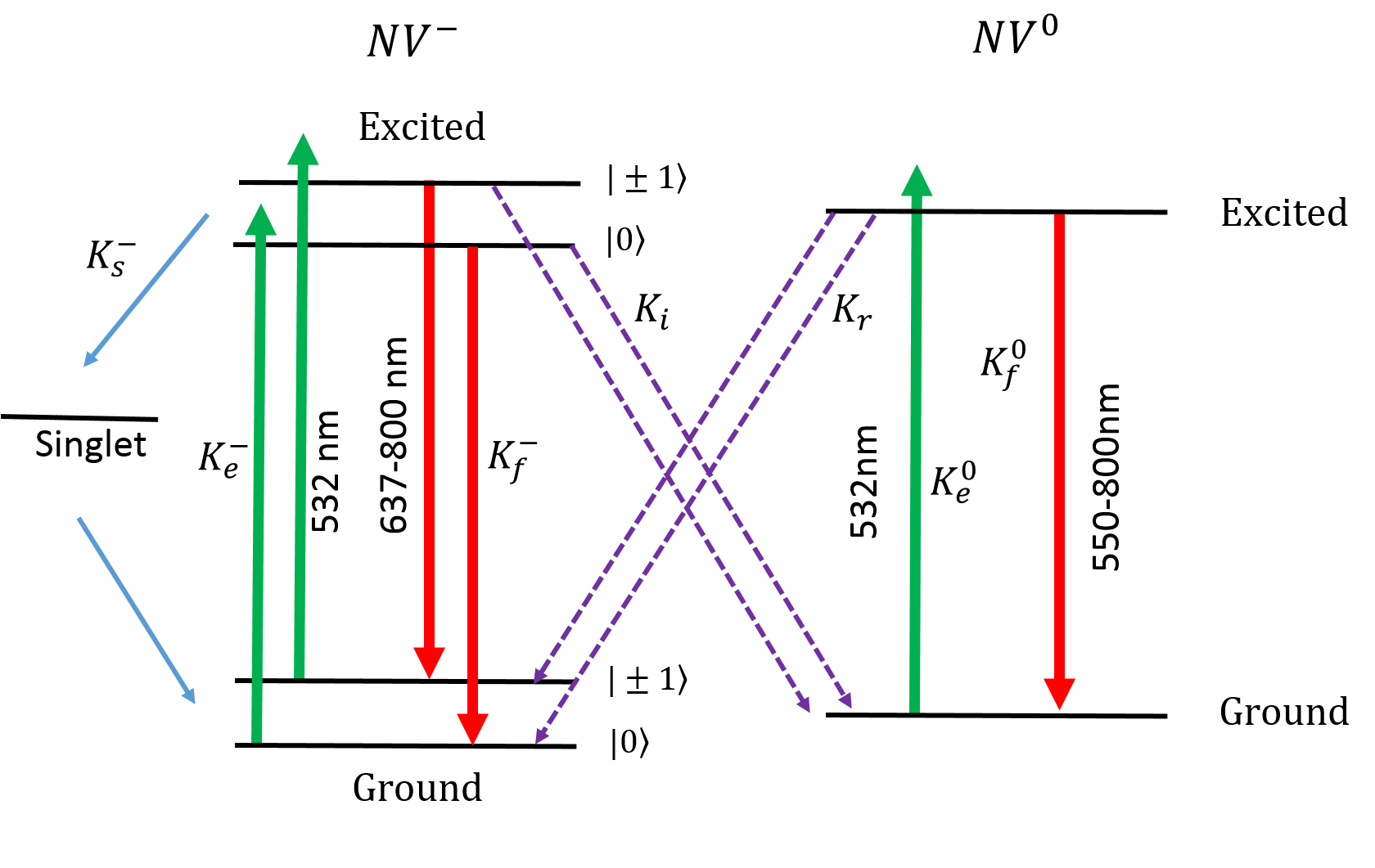}}
\caption{Energy level diagram and relevant transitions for the neutral and negatively charged NV center. The NV$^0$ levels are simplified to include only ground and excited states, and for the NV$^-$ the $m_s=\pm 1$ states are combined, as are the singlet levels, since their detailed inclusion does not modify the analysis presented in this work. $K_e^-$ and $K_e^0$ - green excitation rates (for NV$^-$ and NV$^0$, respectively), depicted by solid green arrows. $K_f^-$ and $K_f^0$ - fluorescence rates, depicted by solid red arrows, $K_s$ - non radiative transition rate, depicted by blue arrow, $K_i$ and $K_r$ - ionization and recombination rates for both green and IR excitations ($K_\alpha = K_{\alpha{_G}} + K_{\alpha_{IR}}$), depicted by dashed purple arrows.} 
\label{fig:energyleveldiagram}

\end{figure}

In order to gain a quantitative understanding of these results, as well as of previously published ones, we consider a rate equation based model for the NV photo-dynamics. 
This simplified model, depicted in Fig. \ref{fig:energyleveldiagram}, consists of the NV$^-$ $m_s = \pm 1$ and $m_s = 0$ spin states in the ground and excited electronic states; the NV$^-$ singlet state; and the NV$^0$ electronic ground and excited states. The considered transitions are shown in the figure, with intra-charge transitions marked by solid arrows, and the I/R transitions by dashed arrows. All transitions are considered spin-independent, except for the transition from the excited state to the singlet state in NV$^-$ \cite{manson_nitrogen-vacancy_2006}. 
 The I/R dynamics used in this model are therefore consistent with the approach suggested in Ref \cite{geiselmann_fast_2013}, although here we did not include additional dark states. We neglect possible contributions of two-photon processes originating from the ground-state to these dynamics, as suggested in Ref. \cite{hopper_near-infrared-assisted_2016} (see also \cite{aslam_photo-induced_2013,hrubesch_efficient_2017}). The internal NV$^-$/NV$^0$ transition rates considered in this model were taken from Ref. \cite{robledo_spin_2011,Storteboom:15} (see \cite{SuppMat}). 

Using this model, we can explain the PL dynamics in Fig. \ref{fig:timeresolved}. The fast quench is caused by the new non-radiative transition from the excited states that is created when we turn on the IR laser. The following slow increase suggests that the combined green and IR induced charge dynamics tend toward enhanced NV$^-$ population compared to the dynamics with green excitation only (even though the steady-state PL under green and IR excitation may be lower than the green only steady-state PL, since PL is not directly related to the NV$^-$ population). The sharp rise as we turn off the IR ($t =$ 60 $\mu$sec) is the opposite process of the sharp quench: By turning off the IR excitation we close the non-radiative transition and recover to the green only rates in the system, with different charge populations (increased NV$^-$ population). Therefore, the fact that the sharp peak is higher than the green only steady-state PL strengthens the claim that the NV$^-$ population increases due to the addition of IR excitation. Lastly, the slow decrease of PL corresponds to the NV$^-$ population slowly decreasing back to its green only steady-state population. 

In contrast to the internal NV$^-$/NV$^0$ rates, the I/R rates from the excited states, for both green and IR light, are unavailable in the literature. 
These single-photon I/R rates can be expressed as $K_{{\alpha}_j}=\frac{\sigma^{\alpha}_j\lambda_j I_j}{hc}$, where $\sigma^{\alpha}_j$ is the cross-section associated with the transition ${\alpha}$ (ionization/recombination) induced by laser $j$, $I_j$ is the intensity of the relevant excitation laser, with $j$ denoting either green (G) or IR.

We first consider the dynamics of the NV$^-$ excited state population ($P^-_e$):
\begin{equation} 
\dot{P}^-_{e} = P^-_g K^-_e-P^-_e (K^-_f+K^-_s+K_{{i}_G}+K_{{i}_{IR}})
\end{equation}

Here $P^-_g$ is the NV$^-$ ground state population. Since the fast quench is approximately two orders of magnitude faster than the slower dynamics (Fig. \ref{fig:timeresolved}. point A), we can assume that the PL level following these fast dynamics represents a quasi-steady-state (QSS) (which does not depend on additional, slower excitation and ionization/recombination processes). We therefore denote the steady-state and QSS populations as $\bar{P}^-_e$ (steady-state with green excitation only, before the quench) and $\bar{P}^-_{e_{IR}}$ (QSS, combined green and IR excitation, following the quench) correspondingly. The fast change in PL is then a result of changing the value of $K_{i_{IR}}$ from zero to some finite value when the IR laser is turned on. For low enough green and IR powers $P^-_g$ remains approximately constant ($\frac{P^-_{g_IR}}{\bar{P}^-_g} \approx 1$). Thus, the dependence of $\bar{P}^-_{e_{IR}}$ on $K_{i_{IR}}$ is given, under the above assumptions, by:
\begin{equation} 
\frac{\bar{P}^-_{e_{IR}}}{\bar{P}^-_e} = \frac{K^-_f+K^-_s+K_{i_G}}{K^-_f+K^-_s+K_{i_{G}}+K_{i_{IR}}},
\label{eq:quenchminus}
\end{equation}
which corresponding to the the fast quench in PL (see \cite{SuppMat}). It is clear from Eq. (\ref{eq:quenchminus}) that the quench will be less significant as the green laser intensity increases, due to the increase in green ionization rate from the excited state (as $K_{i_G} \propto I_G$), while a more significant suppression of PL is expected as the IR laser power increases, due to the added IR ionization rate (again, from the excited state). 

A corresponding analysis and expression can be derived for the normalized fluorescence of NV$^0$:
\begin{equation} 
\frac{\bar{P}^0_{e_{IR}}}{\bar{P}^0_e} = \frac{K_f^0+K_{r_G}}{K_f^0+K_{r_G}+K_{r_{IR}}},
\label{eq:quench0}
\end{equation}

\begin{figure}[tbh]
\begin{center}
\subfigure[]{
\includegraphics[trim = 1mm 1mm 0mm 3mm, clip, width=0.45 \linewidth]{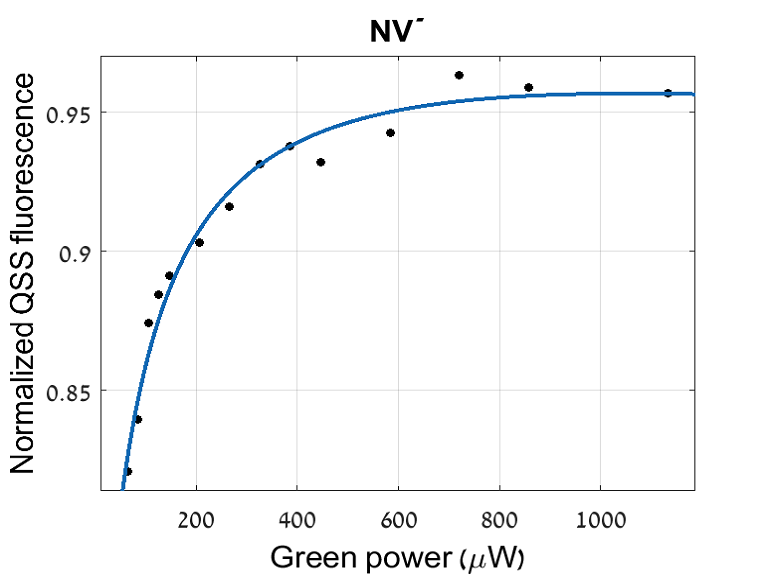}}
\subfigure[]{
\includegraphics[trim = 1mm 1mm 0mm 3mm, clip, width=0.45 \linewidth]{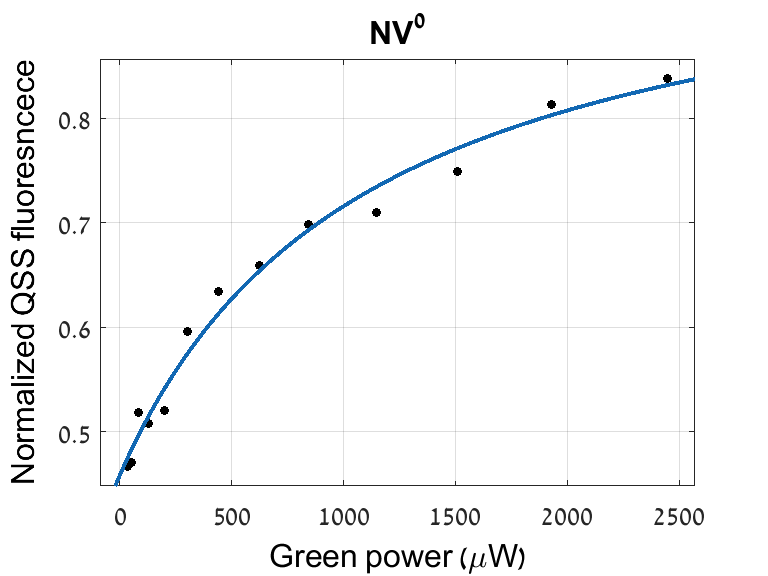}}
\protect\caption {Fast quench of PL dependency on green laser power. QSS PL normalized by SS PL as a function of green laser power for (a) NV$^-$ PL and (b) NV$^0$ PL, with fixed IR power. Black dots - experimental data. Blue curves - fits derived from equations \ref{eq:quenchminus} and \ref{eq:quench0} (see \cite{SuppMat}).}
\label{fig:quenchIR}  
\end{center}
\end{figure}

%
Eqs. \ref{eq:quenchminus} and \ref{eq:quench0} were used to find ionization and recombination rates (as shown in Fig. \ref{fig:quenchIR}) 
through NV$^-$ (a) and NV$^0$ (b) PL. The significant suppression observed indicates that the IR ionization rate from the NV$^-$ excited state is on the order of the excited state decay rate $K_f$. Correspondingly, the strong effect of the green laser power on this quench indicates that the green ionization rate is of the same order. The solid lines in Fig. \ref{fig:quenchIR} are fits to Eq. \ref{eq:quenchminus} and \ref{eq:quench0}, from which we extract the excited state ionization and recombination rates and cross-sections for both green and IR excitations. The uncertainties in the extracted rates stem from the interdependence of the fits on each of the parameters, as well as from the significant differences of the rates between different NVs, which can be caused by small variation in their depths and local environments. 
%
%

The results obtained above from the fast quench in PL at the onset of the IR excitation were used to calculate the full temporal evolution of the PL curves. We further improve the accuracy  of these rates and narrow their uncertainties by applying a numerical optimization algorithm, which fine tunes the values of the cross-sections (within the relevant ranges), such that the model calculation reproduces the experimental data accurately. The results are plotted in Fig. \ref{fig:timeresolved}, represented by a solid red line, 
showing quantitative agreement. 
The cross-sections and their uncertainties, presented in Table \ref{crosssections}, were extracted using this method on several shallow NVs (in the same implanted sample).
We note that the relatively large cross-section for ionization with green laser is surprising, and could shed some light on the spin dynamics of NV centers under green excitation. This rate and its complementary recombination rate can complete and potentially modify the internal NV$^-$ spin dynamics picture, as presented previously  \cite{manson_nitrogen-vacancy_2006,robledo_spin_2011,tetienne_magnetic-field-dependent_2012}. Assuming that the recombination process is not spin dependent (i.e. does not pump preferentially into $m_s=0$), these processes could cause significant spin mixing, even for relatively low laser powers, diminishing laser induced NV$^-$ spin polarization and therefore reducing the efficiency of the NV based measurements. 
An important consequence of the extracted cross-sections is that the ratio between ionization and recombination rates under green illumination results in preferential NV$^0$ population ($P^0_g + P^0_e$) in the steady-state, as opposed to previous results \cite{aslam_photo-induced_2013}. We attribute this discrepancy to the difference between shallow and bulk NVs, and repeat our experimental analysis for bulk NVs. This was done using single NVs in an HPHT sample, using both shallow and bulk NVs (about $2 \mu$m  deep in the same sample). The complete analysis is given \cite{SuppMat}. The shallow NVs resulted in similar cross-sections (within the uncertainty of the measurements) as those of the implanted samples described above. However, for bulk NVs we find that the green ionization rate is reduced by approximately 85\% to $\sigma^i_G = 0.95 \pm 0.47 \cdot 10^{-21}$ m$^2$. The bulk NVs recombination rates were not measured due to low signal and assumed to be the same as for shallow NVs. 
These results highlight the difference in photo-dynamics between bulk and shallow NVs, and are in agreement with previously published data \cite{aslam_photo-induced_2013}. 
%
%

\begin{table}
  \begin{tabular}{|c|c|c|}
  	  \hline
      cross section & value [$m^2$] & rate/power [$\frac{MHz}{mW}$] \\ 
      \hline
      $\sigma^i_G$ & $6.25 \pm 3.12 \cdot 10^{-20}$ & $852 \pm 348$  \\
      $\sigma^i_{IR}$ & $1.76\pm 0.49 \cdot 10^{-22}$ & $1.20 \pm 0.33$ \\
      $\sigma^r_G$ &  $9.83 \pm 4.91 \cdot 10^{-21}$ & $134 \pm 47$ \\
      $\sigma^r_{IR}$  & $4.66 \pm 2.73 \cdot 10^{-22}$ & $3.17 \pm 1.86$ \\
      \hline
  \end{tabular}
  
\caption{Green and IR induced absorption cross sections of the NV$^-$ and NV$^0$ excited states. $\sigma^i_G$ - green induced ionization, $\sigma^i_{IR}$ - IR induced ionization, $\sigma^r_G$ - green induced recombination, $\sigma^r_{IR}$ - IR induced recombination.}
\label{crosssections}
\end{table}

\begin{figure}[tbh]
\begin{center}
\includegraphics[trim = 3mm 1mm 0mm 3mm, clip, width=0.85 \linewidth]{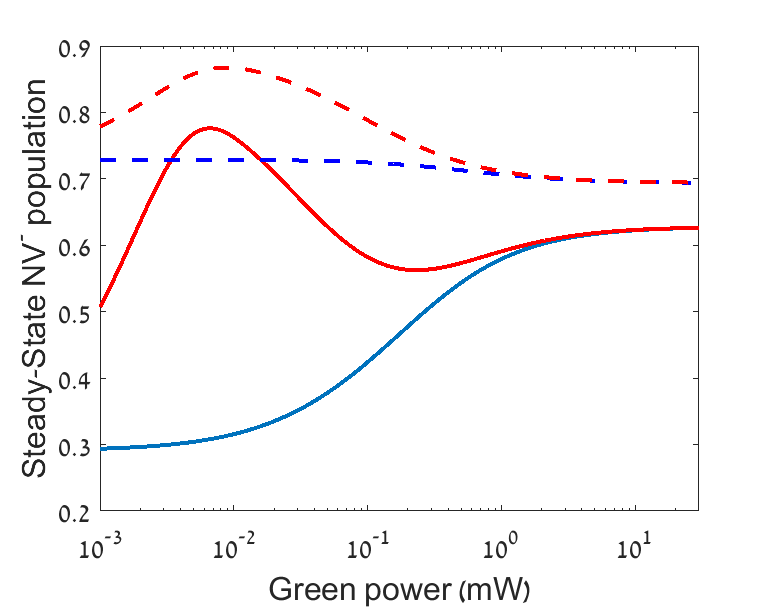}
\protect\caption {IR effect on NV$^-$ steady state population for surface (solid lines) and bulk (dashed lines) NVs. IR excitation power was optimized for strongest effect (25 mW for surface NVs, 35 mW for bulk NVs). Blue - green excitation only. Red - simultaneous excitation with green and IR. Increase of 25\% and 20\% in NV$^-$ population is calculated for surface and bulk NVs respectively, compared to initialization with green only.} 
\label{fig:chargepopulation}
\end{center}
\end{figure}

Finally, in Fig. \ref{fig:chargepopulation} we simulate the steady-state charge-state populations of shallow (solid lines) and bulk (dashed lines) NVs under green (blue lines) or both green and IR illumination (red lines), using the cross-sections extracted above (Table \ref{crosssections}). The values presented in the table are the lower boundaries for the cross sections due to the assumption of no losses in the diamond interface and diffraction limited beam. We find that for shallow NVs, in the presence of only green light, the NVs occupy mostly the unwanted neutral charge state in steady-state for green powers below $220 \mu W$. For higher green excitation powers the negative charge state is preferred, but its occupation saturates at $\sim 63 \% $ NV$^-$. In the presence of both green and IR excitations we see a different trend with a significant increase of the negative charge state for lower powers, followed by a collapse back to the green only excitation steady state at high powers. An increase of $25\%$ of the NV$^-$ population can be achieved by using simultaneous initialization with both green and IR excitations of appropriate intensities, to $\sim 78\%$ population of NV$^-$. 

For bulk NVs, the opposite trend is obtained for this power regime, with the NV$^-$ population decreasing as the green laser power increases, in agreement with the dynamics described in \cite{aslam_photo-induced_2013}. In the presence of both green and IR excitations an increase of the negative charge state is obtained, up to $\sim 87\%$, which corresponds to $\sim 20\%$ increase in the negative charge population.
%
%

In conclusion, we have analyzed the ionization and recombination dynamics of the negatively charged and neutral NV center under green and IR excitation, for a broad range of laser powers. By investigating the features in time-resolved fluorescence measurements for a pulsed excitation sequence, we identified the dominant photo-dynamic processes, and constructed a rate equation model which offers a complete quantitative reproduction of both previously published and present experimental data, some of which were in contradiction \cite{geiselmann_fast_2013,ji_charge_2016}. 
Our dynamic analysis allowed us to directly measure the ionization and recombination cross-sections of the NV$^-$ and NV$^0$ from their respective excited states, under both 532 nm and 1064 nm excitations. Based on these cross-sections we quantitatively identified a significant suppression of NV$^-$ steady-state population under green illumination of shallow NVs as compared to bulk NVs. In addition, these extracted cross-sections allowed us to introduce a method to dramatically increase the NV$^-$ population of shallow NVs, using both green and IR excitation. This result could address various issues encountered empirically for shallow NVs, and offer an approach for improving various relevant applications. 
The results and methods presented here may be used to shed light on some of the fundamental internal NV$^-$ dynamics and rates which are still not fully resolved, e.g. relating to the decay rates from the excited state to the singlet level and from the singlet state to the ground states. Thus, we expect that such insights could play an important role in NV research and potential applications.

\begin{acknowledgments}

\end{acknowledgments}

\bibliography{NV}

\begin{thebibliography}{29}%
\makeatletter
\providecommand \@ifxundefined [1]{%
 \@ifx{#1\undefined}
}%
\providecommand \@ifnum [1]{%
 \ifnum #1\expandafter \@firstoftwo
 \else \expandafter \@secondoftwo
 \fi
}%
\providecommand \@ifx [1]{%
 \ifx #1\expandafter \@firstoftwo
 \else \expandafter \@secondoftwo
 \fi
}%
\providecommand \natexlab [1]{#1}%
\providecommand \enquote  [1]{``#1''}%
\providecommand \bibnamefont  [1]{#1}%
\providecommand \bibfnamefont [1]{#1}%
\providecommand \citenamefont [1]{#1}%
\providecommand \href@noop [0]{\@secondoftwo}%
\providecommand \href [0]{\begingroup \@sanitize@url \@href}%
\providecommand \@href[1]{\@@startlink{#1}\@@href}%
\providecommand \@@href[1]{\endgroup#1\@@endlink}%
\providecommand \@sanitize@url [0]{\catcode `\\12\catcode `\$12\catcode
  `\&12\catcode `\#12\catcode `\^12\catcode `\_12\catcode `\%12\relax}%
\providecommand \@@startlink[1]{}%
\providecommand \@@endlink[0]{}%
\providecommand \url  [0]{\begingroup\@sanitize@url \@url }%
\providecommand \@url [1]{\endgroup\@href {#1}{\urlprefix }}%
\providecommand \urlprefix  [0]{URL }%
\providecommand \Eprint [0]{\href }%
\providecommand \doibase [0]{http://dx.doi.org/}%
\providecommand \selectlanguage [0]{\@gobble}%
\providecommand \bibinfo  [0]{\@secondoftwo}%
\providecommand \bibfield  [0]{\@secondoftwo}%
\providecommand \translation [1]{[#1]}%
\providecommand \BibitemOpen [0]{}%
\providecommand \bibitemStop [0]{}%
\providecommand \bibitemNoStop [0]{.\EOS\space}%
\providecommand \EOS [0]{\spacefactor3000\relax}%
\providecommand \BibitemShut  [1]{\csname bibitem#1\endcsname}%
\let\auto@bib@innerbib\@empty
\bibitem [{\citenamefont {Doherty}\ \emph {et~al.}(2013)\citenamefont
  {Doherty}, \citenamefont {Manson}, \citenamefont {Delaney}, \citenamefont
  {Jelezko}, \citenamefont {Wrachtrup},\ and\ \citenamefont
  {Hollenberg}}]{doherty_nitrogen-vacancy_2013}%
  \BibitemOpen
  \bibfield  {author} {\bibinfo {author} {\bibfnamefont {M.~W.}\ \bibnamefont
  {Doherty}}, \bibinfo {author} {\bibfnamefont {N.~B.}\ \bibnamefont {Manson}},
  \bibinfo {author} {\bibfnamefont {P.}~\bibnamefont {Delaney}}, \bibinfo
  {author} {\bibfnamefont {F.}~\bibnamefont {Jelezko}}, \bibinfo {author}
  {\bibfnamefont {J.}~\bibnamefont {Wrachtrup}}, \ and\ \bibinfo {author}
  {\bibfnamefont {L.~C.}\ \bibnamefont {Hollenberg}},\ }\href {\doibase
  10.1016/j.physrep.2013.02.001} {\bibfield  {journal} {\bibinfo  {journal}
  {Physics Reports}\ }\textbf {\bibinfo {volume} {528}},\ \bibinfo {pages} {1}
  (\bibinfo {year} {2013})}\BibitemShut {NoStop}%
\bibitem [{\citenamefont {Fuchs}\ \emph {et~al.}(2011)\citenamefont {Fuchs},
  \citenamefont {Burkard}, \citenamefont {Klimov},\ and\ \citenamefont
  {Awschalom}}]{Awschalom_memory}%
  \BibitemOpen
  \bibfield  {author} {\bibinfo {author} {\bibfnamefont {G.~D.}\ \bibnamefont
  {Fuchs}}, \bibinfo {author} {\bibfnamefont {G.}~\bibnamefont {Burkard}},
  \bibinfo {author} {\bibfnamefont {P.~V.}\ \bibnamefont {Klimov}}, \ and\
  \bibinfo {author} {\bibfnamefont {D.~D.}\ \bibnamefont {Awschalom}},\ }\href
  {https://search.proquest.com/docview/900176970?accountid=14546} {\bibfield
  {journal} {\bibinfo  {journal} {Nature Physics}\ }\textbf {\bibinfo {volume}
  {7}},\ \bibinfo {pages} {790} (\bibinfo {year} {2011})},\ \bibinfo {note}
  {copyright - Copyright Nature Publishing Group Oct 2011; Last updated -
  2012-11-20}\BibitemShut {NoStop}%
\bibitem [{\citenamefont {Taylor}\ \emph {et~al.}(2008)\citenamefont {Taylor},
  \citenamefont {Cappellaro}, \citenamefont {Childress}, \citenamefont {Jiang},
  \citenamefont {Budker}, \citenamefont {Hemmer}, \citenamefont {Yacoby},
  \citenamefont {Walsworth},\ and\ \citenamefont
  {Lukin}}]{taylor_high-sensitivity_2008}%
  \BibitemOpen
  \bibfield  {author} {\bibinfo {author} {\bibfnamefont {J.~M.}\ \bibnamefont
  {Taylor}}, \bibinfo {author} {\bibfnamefont {P.}~\bibnamefont {Cappellaro}},
  \bibinfo {author} {\bibfnamefont {L.}~\bibnamefont {Childress}}, \bibinfo
  {author} {\bibfnamefont {L.}~\bibnamefont {Jiang}}, \bibinfo {author}
  {\bibfnamefont {D.}~\bibnamefont {Budker}}, \bibinfo {author} {\bibfnamefont
  {P.~R.}\ \bibnamefont {Hemmer}}, \bibinfo {author} {\bibfnamefont
  {A.}~\bibnamefont {Yacoby}}, \bibinfo {author} {\bibfnamefont
  {R.}~\bibnamefont {Walsworth}}, \ and\ \bibinfo {author} {\bibfnamefont
  {M.~D.}\ \bibnamefont {Lukin}},\ }\href {\doibase 10.1038/nphys1075}
  {\bibfield  {journal} {\bibinfo  {journal} {Nature Physics}\ }\textbf
  {\bibinfo {volume} {4}},\ \bibinfo {pages} {810} (\bibinfo {year}
  {2008})}\BibitemShut {NoStop}%
\bibitem [{\citenamefont {Loretz}\ \emph {et~al.}(2014)\citenamefont {Loretz},
  \citenamefont {Pezzagna}, \citenamefont {Meijer},\ and\ \citenamefont
  {Degen}}]{Loretz}%
  \BibitemOpen
  \bibfield  {author} {\bibinfo {author} {\bibfnamefont {M.}~\bibnamefont
  {Loretz}}, \bibinfo {author} {\bibfnamefont {S.}~\bibnamefont {Pezzagna}},
  \bibinfo {author} {\bibfnamefont {J.}~\bibnamefont {Meijer}}, \ and\ \bibinfo
  {author} {\bibfnamefont {C.~L.}\ \bibnamefont {Degen}},\ }\href {\doibase
  10.1063/1.4862749} {\bibfield  {journal} {\bibinfo  {journal} {Applied
  Physics Letters}\ }\textbf {\bibinfo {volume} {104}},\ \bibinfo {pages}
  {033102} (\bibinfo {year} {2014})},\ \Eprint
  {http://arxiv.org/abs/http://dx.doi.org/10.1063/1.4862749}
  {http://dx.doi.org/10.1063/1.4862749} \BibitemShut {NoStop}%
\bibitem [{\citenamefont {Clevenson}\ \emph {et~al.}(2015)\citenamefont
  {Clevenson}, \citenamefont {Trusheim}, \citenamefont {Teale}, \citenamefont
  {Schröder}, \citenamefont {Braje},\ and\ \citenamefont
  {Englund}}]{clevenson_broadband_2015}%
  \BibitemOpen
  \bibfield  {author} {\bibinfo {author} {\bibfnamefont {H.}~\bibnamefont
  {Clevenson}}, \bibinfo {author} {\bibfnamefont {M.~E.}\ \bibnamefont
  {Trusheim}}, \bibinfo {author} {\bibfnamefont {C.}~\bibnamefont {Teale}},
  \bibinfo {author} {\bibfnamefont {T.}~\bibnamefont {Schröder}}, \bibinfo
  {author} {\bibfnamefont {D.}~\bibnamefont {Braje}}, \ and\ \bibinfo {author}
  {\bibfnamefont {D.}~\bibnamefont {Englund}},\ }\href {\doibase
  10.1038/nphys3291} {\bibfield  {journal} {\bibinfo  {journal} {Nature
  Physics}\ }\textbf {\bibinfo {volume} {11}},\ \bibinfo {pages} {393}
  (\bibinfo {year} {2015})}\BibitemShut {NoStop}%
\bibitem [{\citenamefont {Trusheim}\ and\ \citenamefont
  {Englund}(2016)}]{trusheim_wide-field_2016}%
  \BibitemOpen
  \bibfield  {author} {\bibinfo {author} {\bibfnamefont {M.~E.}\ \bibnamefont
  {Trusheim}}\ and\ \bibinfo {author} {\bibfnamefont {D.}~\bibnamefont
  {Englund}},\ }\href {\doibase 10.1088/1367-2630/aa5040} {\bibfield  {journal}
  {\bibinfo  {journal} {New Journal of Physics}\ }\textbf {\bibinfo {volume}
  {18}},\ \bibinfo {pages} {123023} (\bibinfo {year} {2016})}\BibitemShut
  {NoStop}%
\bibitem [{\citenamefont {Dolde}\ \emph {et~al.}(2011)\citenamefont {Dolde},
  \citenamefont {Fedder}, \citenamefont {Doherty}, \citenamefont {Nöbauer},
  \citenamefont {Rempp}, \citenamefont {Balasubramanian}, \citenamefont {Wolf},
  \citenamefont {Reinhard}, \citenamefont {Hollenberg}, \citenamefont
  {Jelezko},\ and\ \citenamefont {Wrachtrup}}]{dolde_electric-field_2011}%
  \BibitemOpen
  \bibfield  {author} {\bibinfo {author} {\bibfnamefont {F.}~\bibnamefont
  {Dolde}}, \bibinfo {author} {\bibfnamefont {H.}~\bibnamefont {Fedder}},
  \bibinfo {author} {\bibfnamefont {M.~W.}\ \bibnamefont {Doherty}}, \bibinfo
  {author} {\bibfnamefont {T.}~\bibnamefont {Nöbauer}}, \bibinfo {author}
  {\bibfnamefont {F.}~\bibnamefont {Rempp}}, \bibinfo {author} {\bibfnamefont
  {G.}~\bibnamefont {Balasubramanian}}, \bibinfo {author} {\bibfnamefont
  {T.}~\bibnamefont {Wolf}}, \bibinfo {author} {\bibfnamefont {F.}~\bibnamefont
  {Reinhard}}, \bibinfo {author} {\bibfnamefont {L.~C.~L.}\ \bibnamefont
  {Hollenberg}}, \bibinfo {author} {\bibfnamefont {F.}~\bibnamefont {Jelezko}},
  \ and\ \bibinfo {author} {\bibfnamefont {J.}~\bibnamefont {Wrachtrup}},\
  }\href {\doibase 10.1038/nphys1969} {\bibfield  {journal} {\bibinfo
  {journal} {Nature Physics}\ }\textbf {\bibinfo {volume} {7}},\ \bibinfo
  {pages} {459} (\bibinfo {year} {2011})}\BibitemShut {NoStop}%
\bibitem [{\citenamefont {Staudacher}\ \emph {et~al.}(2013)\citenamefont
  {Staudacher}, \citenamefont {Shi}, \citenamefont {Pezzagna}, \citenamefont
  {Meijer}, \citenamefont {Du}, \citenamefont {Meriles}, \citenamefont
  {Reinhard},\ and\ \citenamefont {Wrachtrup}}]{staudacher_nuclear_2013}%
  \BibitemOpen
  \bibfield  {author} {\bibinfo {author} {\bibfnamefont {T.}~\bibnamefont
  {Staudacher}}, \bibinfo {author} {\bibfnamefont {F.}~\bibnamefont {Shi}},
  \bibinfo {author} {\bibfnamefont {S.}~\bibnamefont {Pezzagna}}, \bibinfo
  {author} {\bibfnamefont {J.}~\bibnamefont {Meijer}}, \bibinfo {author}
  {\bibfnamefont {J.}~\bibnamefont {Du}}, \bibinfo {author} {\bibfnamefont
  {C.~A.}\ \bibnamefont {Meriles}}, \bibinfo {author} {\bibfnamefont
  {F.}~\bibnamefont {Reinhard}}, \ and\ \bibinfo {author} {\bibfnamefont
  {J.}~\bibnamefont {Wrachtrup}},\ }\href {\doibase 10.1126/science.1231675}
  {\bibfield  {journal} {\bibinfo  {journal} {Science}\ }\textbf {\bibinfo
  {volume} {339}},\ \bibinfo {pages} {561} (\bibinfo {year}
  {2013})}\BibitemShut {NoStop}%
\bibitem [{\citenamefont {Mamin}\ \emph {et~al.}(2013)\citenamefont {Mamin},
  \citenamefont {Kim}, \citenamefont {Sherwood}, \citenamefont {Rettner},
  \citenamefont {Ohno}, \citenamefont {Awschalom},\ and\ \citenamefont
  {Rugar}}]{mamin_nanoscale_2013}%
  \BibitemOpen
  \bibfield  {author} {\bibinfo {author} {\bibfnamefont {H.~J.}\ \bibnamefont
  {Mamin}}, \bibinfo {author} {\bibfnamefont {M.}~\bibnamefont {Kim}}, \bibinfo
  {author} {\bibfnamefont {M.~H.}\ \bibnamefont {Sherwood}}, \bibinfo {author}
  {\bibfnamefont {C.~T.}\ \bibnamefont {Rettner}}, \bibinfo {author}
  {\bibfnamefont {K.}~\bibnamefont {Ohno}}, \bibinfo {author} {\bibfnamefont
  {D.~D.}\ \bibnamefont {Awschalom}}, \ and\ \bibinfo {author} {\bibfnamefont
  {D.}~\bibnamefont {Rugar}},\ }\href {\doibase 10.1126/science.1231540}
  {\bibfield  {journal} {\bibinfo  {journal} {Science}\ }\textbf {\bibinfo
  {volume} {339}},\ \bibinfo {pages} {557} (\bibinfo {year}
  {2013})}\BibitemShut {NoStop}%
\bibitem [{\citenamefont {Bar-Gill}\ \emph {et~al.}(2013)\citenamefont
  {Bar-Gill}, \citenamefont {Pham}, \citenamefont {Jarmola}, \citenamefont
  {Budker},\ and\ \citenamefont {Walsworth}}]{bar-gill_solid-state_2013}%
  \BibitemOpen
  \bibfield  {author} {\bibinfo {author} {\bibfnamefont {N.}~\bibnamefont
  {Bar-Gill}}, \bibinfo {author} {\bibfnamefont {L.}~\bibnamefont {Pham}},
  \bibinfo {author} {\bibfnamefont {A.}~\bibnamefont {Jarmola}}, \bibinfo
  {author} {\bibfnamefont {D.}~\bibnamefont {Budker}}, \ and\ \bibinfo {author}
  {\bibfnamefont {R.}~\bibnamefont {Walsworth}},\ }\href {\doibase
  10.1038/ncomms2771} {\bibfield  {journal} {\bibinfo  {journal} {Nature
  Communications}\ }\textbf {\bibinfo {volume} {4}},\ \bibinfo {pages} {1743}
  (\bibinfo {year} {2013})}\BibitemShut {NoStop}%
\bibitem [{\citenamefont {Farfurnik}\ \emph {et~al.}(2015)\citenamefont
  {Farfurnik}, \citenamefont {Jarmola}, \citenamefont {Pham}, \citenamefont
  {Wang}, \citenamefont {Dobrovitski}, \citenamefont {Walsworth}, \citenamefont
  {Budker},\ and\ \citenamefont {Bar-Gill}}]{farfurnik_optimizing_2015}%
  \BibitemOpen
  \bibfield  {author} {\bibinfo {author} {\bibfnamefont {D.}~\bibnamefont
  {Farfurnik}}, \bibinfo {author} {\bibfnamefont {A.}~\bibnamefont {Jarmola}},
  \bibinfo {author} {\bibfnamefont {L.~M.}\ \bibnamefont {Pham}}, \bibinfo
  {author} {\bibfnamefont {Z.~H.}\ \bibnamefont {Wang}}, \bibinfo {author}
  {\bibfnamefont {V.~V.}\ \bibnamefont {Dobrovitski}}, \bibinfo {author}
  {\bibfnamefont {R.~L.}\ \bibnamefont {Walsworth}}, \bibinfo {author}
  {\bibfnamefont {D.}~\bibnamefont {Budker}}, \ and\ \bibinfo {author}
  {\bibfnamefont {N.}~\bibnamefont {Bar-Gill}},\ }\href {\doibase
  10.1103/PhysRevB.92.060301} {\bibfield  {journal} {\bibinfo  {journal}
  {Physical Review B}\ }\textbf {\bibinfo {volume} {92}} (\bibinfo {year}
  {2015}),\ 10.1103/PhysRevB.92.060301}\BibitemShut {NoStop}%
\bibitem [{\citenamefont {Childress}\ \emph {et~al.}(2006)\citenamefont
  {Childress}, \citenamefont {Gurudev~Dutt}, \citenamefont {Taylor},
  \citenamefont {Zibrov}, \citenamefont {Jelezko}, \citenamefont {Wrachtrup},
  \citenamefont {Hemmer},\ and\ \citenamefont
  {Lukin}}]{childress_coherent_2006}%
  \BibitemOpen
  \bibfield  {author} {\bibinfo {author} {\bibfnamefont {L.}~\bibnamefont
  {Childress}}, \bibinfo {author} {\bibfnamefont {M.~V.}\ \bibnamefont
  {Gurudev~Dutt}}, \bibinfo {author} {\bibfnamefont {J.~M.}\ \bibnamefont
  {Taylor}}, \bibinfo {author} {\bibfnamefont {A.~S.}\ \bibnamefont {Zibrov}},
  \bibinfo {author} {\bibfnamefont {F.}~\bibnamefont {Jelezko}}, \bibinfo
  {author} {\bibfnamefont {J.}~\bibnamefont {Wrachtrup}}, \bibinfo {author}
  {\bibfnamefont {P.~R.}\ \bibnamefont {Hemmer}}, \ and\ \bibinfo {author}
  {\bibfnamefont {M.~D.}\ \bibnamefont {Lukin}},\ }\href {\doibase
  10.1126/science.1131871} {\bibfield  {journal} {\bibinfo  {journal}
  {Science}\ }\textbf {\bibinfo {volume} {314}},\ \bibinfo {pages} {281}
  (\bibinfo {year} {2006})}\BibitemShut {NoStop}%
\bibitem [{\citenamefont {Robledo}\ \emph
  {et~al.}(2011{\natexlab{a}})\citenamefont {Robledo}, \citenamefont
  {Childress}, \citenamefont {Bernien}, \citenamefont {Hensen}, \citenamefont
  {Alkemade},\ and\ \citenamefont {Hanson}}]{robledo_high-fidelity_2011}%
  \BibitemOpen
  \bibfield  {author} {\bibinfo {author} {\bibfnamefont {L.}~\bibnamefont
  {Robledo}}, \bibinfo {author} {\bibfnamefont {L.}~\bibnamefont {Childress}},
  \bibinfo {author} {\bibfnamefont {H.}~\bibnamefont {Bernien}}, \bibinfo
  {author} {\bibfnamefont {B.}~\bibnamefont {Hensen}}, \bibinfo {author}
  {\bibfnamefont {P.~F.~A.}\ \bibnamefont {Alkemade}}, \ and\ \bibinfo {author}
  {\bibfnamefont {R.}~\bibnamefont {Hanson}},\ }\href {\doibase
  10.1038/nature10401} {\bibfield  {journal} {\bibinfo  {journal} {Nature}\
  }\textbf {\bibinfo {volume} {477}},\ \bibinfo {pages} {574} (\bibinfo {year}
  {2011}{\natexlab{a}})}\BibitemShut {NoStop}%
\bibitem [{\citenamefont {Romach}\ \emph {et~al.}(2015)\citenamefont {Romach},
  \citenamefont {Müller}, \citenamefont {Unden}, \citenamefont {Rogers},
  \citenamefont {Isoda}, \citenamefont {Itoh}, \citenamefont {Markham},
  \citenamefont {Stacey}, \citenamefont {Meijer}, \citenamefont {Pezzagna},
  \citenamefont {Naydenov}, \citenamefont {McGuinness}, \citenamefont
  {Bar-Gill},\ and\ \citenamefont {Jelezko}}]{romach_spectroscopy_2015}%
  \BibitemOpen
  \bibfield  {author} {\bibinfo {author} {\bibfnamefont {Y.}~\bibnamefont
  {Romach}}, \bibinfo {author} {\bibfnamefont {C.}~\bibnamefont {Müller}},
  \bibinfo {author} {\bibfnamefont {T.}~\bibnamefont {Unden}}, \bibinfo
  {author} {\bibfnamefont {L.}~\bibnamefont {Rogers}}, \bibinfo {author}
  {\bibfnamefont {T.}~\bibnamefont {Isoda}}, \bibinfo {author} {\bibfnamefont
  {K.}~\bibnamefont {Itoh}}, \bibinfo {author} {\bibfnamefont {M.}~\bibnamefont
  {Markham}}, \bibinfo {author} {\bibfnamefont {A.}~\bibnamefont {Stacey}},
  \bibinfo {author} {\bibfnamefont {J.}~\bibnamefont {Meijer}}, \bibinfo
  {author} {\bibfnamefont {S.}~\bibnamefont {Pezzagna}}, \bibinfo {author}
  {\bibfnamefont {B.}~\bibnamefont {Naydenov}}, \bibinfo {author}
  {\bibfnamefont {L.}~\bibnamefont {McGuinness}}, \bibinfo {author}
  {\bibfnamefont {N.}~\bibnamefont {Bar-Gill}}, \ and\ \bibinfo {author}
  {\bibfnamefont {F.}~\bibnamefont {Jelezko}},\ }\href {\doibase
  10.1103/PhysRevLett.114.017601} {\bibfield  {journal} {\bibinfo  {journal}
  {Physical Review Letters}\ }\textbf {\bibinfo {volume} {114}} (\bibinfo
  {year} {2015}),\ 10.1103/PhysRevLett.114.017601}\BibitemShut {NoStop}%
\bibitem [{\citenamefont {Ohno}\ \emph {et~al.}(2012)\citenamefont {Ohno},
  \citenamefont {Joseph~Heremans}, \citenamefont {Bassett}, \citenamefont
  {Myers}, \citenamefont {Toyli}, \citenamefont {Bleszynski~Jayich},
  \citenamefont {Palmstrøm},\ and\ \citenamefont
  {Awschalom}}]{ohno_engineering_2012}%
  \BibitemOpen
  \bibfield  {author} {\bibinfo {author} {\bibfnamefont {K.}~\bibnamefont
  {Ohno}}, \bibinfo {author} {\bibfnamefont {F.}~\bibnamefont
  {Joseph~Heremans}}, \bibinfo {author} {\bibfnamefont {L.~C.}\ \bibnamefont
  {Bassett}}, \bibinfo {author} {\bibfnamefont {B.~A.}\ \bibnamefont {Myers}},
  \bibinfo {author} {\bibfnamefont {D.~M.}\ \bibnamefont {Toyli}}, \bibinfo
  {author} {\bibfnamefont {A.~C.}\ \bibnamefont {Bleszynski~Jayich}}, \bibinfo
  {author} {\bibfnamefont {C.~J.}\ \bibnamefont {Palmstrøm}}, \ and\ \bibinfo
  {author} {\bibfnamefont {D.~D.}\ \bibnamefont {Awschalom}},\ }\href {\doibase
  10.1063/1.4748280} {\bibfield  {journal} {\bibinfo  {journal} {Applied
  Physics Letters}\ }\textbf {\bibinfo {volume} {101}},\ \bibinfo {pages}
  {082413} (\bibinfo {year} {2012})}\BibitemShut {NoStop}%
\bibitem [{\citenamefont {Wang}\ \emph {et~al.}(2016)\citenamefont {Wang},
  \citenamefont {Zhang}, \citenamefont {Zhang}, \citenamefont {You},
  \citenamefont {Li}, \citenamefont {Guo}, \citenamefont {Feng}, \citenamefont
  {Song}, \citenamefont {Lou}, \citenamefont {Zhu},\ and\ \citenamefont
  {Wang}}]{WangCoherence2016}%
  \BibitemOpen
  \bibfield  {author} {\bibinfo {author} {\bibfnamefont {J.}~\bibnamefont
  {Wang}}, \bibinfo {author} {\bibfnamefont {W.}~\bibnamefont {Zhang}},
  \bibinfo {author} {\bibfnamefont {J.}~\bibnamefont {Zhang}}, \bibinfo
  {author} {\bibfnamefont {J.}~\bibnamefont {You}}, \bibinfo {author}
  {\bibfnamefont {Y.}~\bibnamefont {Li}}, \bibinfo {author} {\bibfnamefont
  {G.}~\bibnamefont {Guo}}, \bibinfo {author} {\bibfnamefont {F.}~\bibnamefont
  {Feng}}, \bibinfo {author} {\bibfnamefont {X.}~\bibnamefont {Song}}, \bibinfo
  {author} {\bibfnamefont {L.}~\bibnamefont {Lou}}, \bibinfo {author}
  {\bibfnamefont {W.}~\bibnamefont {Zhu}}, \ and\ \bibinfo {author}
  {\bibfnamefont {G.}~\bibnamefont {Wang}},\ }\href {\doibase
  10.1039/C5NR08690F} {\bibfield  {journal} {\bibinfo  {journal} {Nanoscale}\
  }\textbf {\bibinfo {volume} {8}},\ \bibinfo {pages} {5780} (\bibinfo {year}
  {2016})}\BibitemShut {NoStop}%
\bibitem [{\citenamefont {Shields}\ \emph {et~al.}(2015)\citenamefont
  {Shields}, \citenamefont {Unterreithmeier}, \citenamefont {de~Leon},
  \citenamefont {Park},\ and\ \citenamefont
  {Lukin}}]{Shields_ChargeStateReadout_2015}%
  \BibitemOpen
  \bibfield  {author} {\bibinfo {author} {\bibfnamefont {B.~J.}\ \bibnamefont
  {Shields}}, \bibinfo {author} {\bibfnamefont {Q.~P.}\ \bibnamefont
  {Unterreithmeier}}, \bibinfo {author} {\bibfnamefont {N.~P.}\ \bibnamefont
  {de~Leon}}, \bibinfo {author} {\bibfnamefont {H.}~\bibnamefont {Park}}, \
  and\ \bibinfo {author} {\bibfnamefont {M.~D.}\ \bibnamefont {Lukin}},\ }\href
  {\doibase 10.1103/PhysRevLett.114.136402} {\bibfield  {journal} {\bibinfo
  {journal} {Phys. Rev. Lett.}\ }\textbf {\bibinfo {volume} {114}},\ \bibinfo
  {pages} {136402} (\bibinfo {year} {2015})}\BibitemShut {NoStop}%
\bibitem [{\citenamefont {Jayakumar}\ \emph {et~al.}(2016)\citenamefont
  {Jayakumar}, \citenamefont {Henshaw}, \citenamefont {Dhomkar}, \citenamefont
  {Pagliero}, \citenamefont {Laraoui}, \citenamefont {Manson}, \citenamefont
  {Albu}, \citenamefont {Doherty},\ and\ \citenamefont
  {Meriles}}]{jayakumar_OpticalPatterning_2016}%
  \BibitemOpen
  \bibfield  {author} {\bibinfo {author} {\bibfnamefont {H.}~\bibnamefont
  {Jayakumar}}, \bibinfo {author} {\bibfnamefont {J.}~\bibnamefont {Henshaw}},
  \bibinfo {author} {\bibfnamefont {S.}~\bibnamefont {Dhomkar}}, \bibinfo
  {author} {\bibfnamefont {D.}~\bibnamefont {Pagliero}}, \bibinfo {author}
  {\bibfnamefont {A.}~\bibnamefont {Laraoui}}, \bibinfo {author} {\bibfnamefont
  {N.~B.}\ \bibnamefont {Manson}}, \bibinfo {author} {\bibfnamefont
  {R.}~\bibnamefont {Albu}}, \bibinfo {author} {\bibfnamefont {M.~W.}\
  \bibnamefont {Doherty}}, \ and\ \bibinfo {author} {\bibfnamefont {C.~A.}\
  \bibnamefont {Meriles}},\ }\href {\doibase 10.1038/ncomms12660} {\bibfield
  {journal} {\bibinfo  {journal} {Nature Communications}\ }\textbf {\bibinfo
  {volume} {7}},\ \bibinfo {pages} {12660} (\bibinfo {year}
  {2016})}\BibitemShut {NoStop}%
\bibitem [{\citenamefont {Geiselmann}\ \emph {et~al.}(2013)\citenamefont
  {Geiselmann}, \citenamefont {Marty}, \citenamefont {Garc?a~de Abajo},\ and\
  \citenamefont {Quidant}}]{geiselmann_fast_2013}%
  \BibitemOpen
  \bibfield  {author} {\bibinfo {author} {\bibfnamefont {M.}~\bibnamefont
  {Geiselmann}}, \bibinfo {author} {\bibfnamefont {R.}~\bibnamefont {Marty}},
  \bibinfo {author} {\bibfnamefont {F.~J.}\ \bibnamefont {Garc?a~de Abajo}}, \
  and\ \bibinfo {author} {\bibfnamefont {R.}~\bibnamefont {Quidant}},\ }\href
  {\doibase 10.1038/nphys2770} {\bibfield  {journal} {\bibinfo  {journal}
  {Nature Physics}\ }\textbf {\bibinfo {volume} {9}},\ \bibinfo {pages} {785}
  (\bibinfo {year} {2013})}\BibitemShut {NoStop}%
\bibitem [{\citenamefont {Hopper}\ \emph {et~al.}(2016)\citenamefont {Hopper},
  \citenamefont {Grote}, \citenamefont {Exarhos},\ and\ \citenamefont
  {Bassett}}]{hopper_near-infrared-assisted_2016}%
  \BibitemOpen
  \bibfield  {author} {\bibinfo {author} {\bibfnamefont {D.~A.}\ \bibnamefont
  {Hopper}}, \bibinfo {author} {\bibfnamefont {R.~R.}\ \bibnamefont {Grote}},
  \bibinfo {author} {\bibfnamefont {A.~L.}\ \bibnamefont {Exarhos}}, \ and\
  \bibinfo {author} {\bibfnamefont {L.~C.}\ \bibnamefont {Bassett}},\ }\href
  {\doibase 10.1103/PhysRevB.94.241201} {\bibfield  {journal} {\bibinfo
  {journal} {Physical Review B}\ }\textbf {\bibinfo {volume} {94}} (\bibinfo
  {year} {2016}),\ 10.1103/PhysRevB.94.241201}\BibitemShut {NoStop}%
\bibitem [{\citenamefont {Lai}\ \emph {et~al.}(2013)\citenamefont {Lai},
  \citenamefont {Faklaris}, \citenamefont {Zheng}, \citenamefont {Jacques},
  \citenamefont {Chang}, \citenamefont {Roch},\ and\ \citenamefont
  {Treussart}}]{Lai_2013}%
  \BibitemOpen
  \bibfield  {author} {\bibinfo {author} {\bibfnamefont {N.~D.}\ \bibnamefont
  {Lai}}, \bibinfo {author} {\bibfnamefont {O.}~\bibnamefont {Faklaris}},
  \bibinfo {author} {\bibfnamefont {D.}~\bibnamefont {Zheng}}, \bibinfo
  {author} {\bibfnamefont {V.}~\bibnamefont {Jacques}}, \bibinfo {author}
  {\bibfnamefont {H.-C.}\ \bibnamefont {Chang}}, \bibinfo {author}
  {\bibfnamefont {J.-F.}\ \bibnamefont {Roch}}, \ and\ \bibinfo {author}
  {\bibfnamefont {F.}~\bibnamefont {Treussart}},\ }\href
  {http://stacks.iop.org/1367-2630/15/i=3/a=033030} {\bibfield  {journal}
  {\bibinfo  {journal} {New Journal of Physics}\ }\textbf {\bibinfo {volume}
  {15}},\ \bibinfo {pages} {033030} (\bibinfo {year} {2013})}\BibitemShut
  {NoStop}%
\bibitem [{\citenamefont {Ji}\ and\ \citenamefont
  {Dutt}(2016)}]{ji_charge_2016}%
  \BibitemOpen
  \bibfield  {author} {\bibinfo {author} {\bibfnamefont {P.}~\bibnamefont
  {Ji}}\ and\ \bibinfo {author} {\bibfnamefont {M.~V.~G.}\ \bibnamefont
  {Dutt}},\ }\href {\doibase 10.1103/PhysRevB.94.024101} {\bibfield  {journal}
  {\bibinfo  {journal} {Physical Review B}\ }\textbf {\bibinfo {volume} {94}}
  (\bibinfo {year} {2016}),\ 10.1103/PhysRevB.94.024101}\BibitemShut {NoStop}%
\bibitem [{\citenamefont {Aslam}\ \emph {et~al.}(2013)\citenamefont {Aslam},
  \citenamefont {Waldherr}, \citenamefont {Neumann}, \citenamefont {Jelezko},\
  and\ \citenamefont {Wrachtrup}}]{aslam_photo-induced_2013}%
  \BibitemOpen
  \bibfield  {author} {\bibinfo {author} {\bibfnamefont {N.}~\bibnamefont
  {Aslam}}, \bibinfo {author} {\bibfnamefont {G.}~\bibnamefont {Waldherr}},
  \bibinfo {author} {\bibfnamefont {P.}~\bibnamefont {Neumann}}, \bibinfo
  {author} {\bibfnamefont {F.}~\bibnamefont {Jelezko}}, \ and\ \bibinfo
  {author} {\bibfnamefont {J.}~\bibnamefont {Wrachtrup}},\ }\href {\doibase
  10.1088/1367-2630/15/1/013064} {\bibfield  {journal} {\bibinfo  {journal}
  {New Journal of Physics}\ }\textbf {\bibinfo {volume} {15}},\ \bibinfo
  {pages} {013064} (\bibinfo {year} {2013})}\BibitemShut {NoStop}%
\bibitem [{\citenamefont {Manson}\ \emph {et~al.}(2006)\citenamefont {Manson},
  \citenamefont {Harrison},\ and\ \citenamefont
  {Sellars}}]{manson_nitrogen-vacancy_2006}%
  \BibitemOpen
  \bibfield  {author} {\bibinfo {author} {\bibfnamefont {N.~B.}\ \bibnamefont
  {Manson}}, \bibinfo {author} {\bibfnamefont {J.~P.}\ \bibnamefont
  {Harrison}}, \ and\ \bibinfo {author} {\bibfnamefont {M.~J.}\ \bibnamefont
  {Sellars}},\ }\href {\doibase 10.1103/PhysRevB.74.104303} {\bibfield
  {journal} {\bibinfo  {journal} {Physical Review B}\ }\textbf {\bibinfo
  {volume} {74}} (\bibinfo {year} {2006}),\
  10.1103/PhysRevB.74.104303}\BibitemShut {NoStop}%
\bibitem [{\citenamefont {Robledo}\ \emph
  {et~al.}(2011{\natexlab{b}})\citenamefont {Robledo}, \citenamefont {Bernien},
  \citenamefont {Sar},\ and\ \citenamefont {Hanson}}]{robledo_spin_2011}%
  \BibitemOpen
  \bibfield  {author} {\bibinfo {author} {\bibfnamefont {L.}~\bibnamefont
  {Robledo}}, \bibinfo {author} {\bibfnamefont {H.}~\bibnamefont {Bernien}},
  \bibinfo {author} {\bibfnamefont {T.~v.~d.}\ \bibnamefont {Sar}}, \ and\
  \bibinfo {author} {\bibfnamefont {R.}~\bibnamefont {Hanson}},\ }\href
  {\doibase 10.1088/1367-2630/13/2/025013} {\bibfield  {journal} {\bibinfo
  {journal} {New Journal of Physics}\ }\textbf {\bibinfo {volume} {13}},\
  \bibinfo {pages} {025013} (\bibinfo {year} {2011}{\natexlab{b}})}\BibitemShut
  {NoStop}%
\bibitem [{\citenamefont {I~Meirzada}\ and\ \citenamefont
  {Bar-Gill}()}]{SuppMat}%
  \BibitemOpen
  \bibfield  {author} {\bibinfo {author} {\bibfnamefont {S.~A.~W.}\
  \bibnamefont {I~Meirzada}, \bibfnamefont {Y~Hovav}}\ and\ \bibinfo {author}
  {\bibfnamefont {N.}~\bibnamefont {Bar-Gill}},\ }\href@noop {} {\enquote
  {\bibinfo {title} {Negative charge enhancement - supplementary
  information},}\ }\bibinfo {note} {See Supplemental Material}\BibitemShut
  {NoStop}%
\bibitem [{\citenamefont {Hrubesch}\ \emph {et~al.}(2017)\citenamefont
  {Hrubesch}, \citenamefont {Braunbeck}, \citenamefont {Stutzmann},
  \citenamefont {Reinhard},\ and\ \citenamefont
  {Brandt}}]{hrubesch_efficient_2017}%
  \BibitemOpen
  \bibfield  {author} {\bibinfo {author} {\bibfnamefont {F.~M.}\ \bibnamefont
  {Hrubesch}}, \bibinfo {author} {\bibfnamefont {G.}~\bibnamefont {Braunbeck}},
  \bibinfo {author} {\bibfnamefont {M.}~\bibnamefont {Stutzmann}}, \bibinfo
  {author} {\bibfnamefont {F.}~\bibnamefont {Reinhard}}, \ and\ \bibinfo
  {author} {\bibfnamefont {M.~S.}\ \bibnamefont {Brandt}},\ }\href {\doibase
  10.1103/PhysRevLett.118.037601} {\bibfield  {journal} {\bibinfo  {journal}
  {Physical Review Letters}\ }\textbf {\bibinfo {volume} {118}} (\bibinfo
  {year} {2017}),\ 10.1103/PhysRevLett.118.037601}\BibitemShut {NoStop}%
\bibitem [{\citenamefont {Storteboom}\ \emph {et~al.}(2015)\citenamefont
  {Storteboom}, \citenamefont {Dolan}, \citenamefont {Castelletto},
  \citenamefont {Li},\ and\ \citenamefont {Gu}}]{Storteboom:15}%
  \BibitemOpen
  \bibfield  {author} {\bibinfo {author} {\bibfnamefont {J.}~\bibnamefont
  {Storteboom}}, \bibinfo {author} {\bibfnamefont {P.}~\bibnamefont {Dolan}},
  \bibinfo {author} {\bibfnamefont {S.}~\bibnamefont {Castelletto}}, \bibinfo
  {author} {\bibfnamefont {X.}~\bibnamefont {Li}}, \ and\ \bibinfo {author}
  {\bibfnamefont {M.}~\bibnamefont {Gu}},\ }\href {\doibase
  10.1364/OE.23.011327} {\bibfield  {journal} {\bibinfo  {journal} {Opt.
  Express}\ }\textbf {\bibinfo {volume} {23}},\ \bibinfo {pages} {11327}
  (\bibinfo {year} {2015})}\BibitemShut {NoStop}%
\bibitem [{\citenamefont {Tetienne}\ \emph {et~al.}(2012)\citenamefont
  {Tetienne}, \citenamefont {Rondin}, \citenamefont {Spinicelli}, \citenamefont
  {Chipaux}, \citenamefont {Debuisschert}, \citenamefont {Roch},\ and\
  \citenamefont {Jacques}}]{tetienne_magnetic-field-dependent_2012}%
  \BibitemOpen
  \bibfield  {author} {\bibinfo {author} {\bibfnamefont {J.-P.}\ \bibnamefont
  {Tetienne}}, \bibinfo {author} {\bibfnamefont {L.}~\bibnamefont {Rondin}},
  \bibinfo {author} {\bibfnamefont {P.}~\bibnamefont {Spinicelli}}, \bibinfo
  {author} {\bibfnamefont {M.}~\bibnamefont {Chipaux}}, \bibinfo {author}
  {\bibfnamefont {T.}~\bibnamefont {Debuisschert}}, \bibinfo {author}
  {\bibfnamefont {J.-F.}\ \bibnamefont {Roch}}, \ and\ \bibinfo {author}
  {\bibfnamefont {V.}~\bibnamefont {Jacques}},\ }\href {\doibase
  10.1088/1367-2630/14/10/103033} {\bibfield  {journal} {\bibinfo  {journal}
  {New Journal of Physics}\ }\textbf {\bibinfo {volume} {14}},\ \bibinfo
  {pages} {103033} (\bibinfo {year} {2012})}\BibitemShut {NoStop}%
\end{thebibliography}%

\end{document}